# Silicon nitride PIC beam formers for light sheet fluorescent microscopy

ALIREZA TABATABAEI MASHAYEKH[1,*] AND JEREMY WITZENS[1]

[1]*Institute of Integrated Photonics, RWTH Aachen University, Campus Boulevard 73, 52074 Aachen, Germany*
*\*amashayekh@iph.rwth-aachen.de*

**Abstract:** Light sheet fluorescence microscopy (LSFM) has transformed the way we visualize biological tissues in three dimensions, offering high-resolution imaging while minimizing photo-induced damage to the samples. Recent breakthroughs in tissue-clearing methods have further improved LSFM's capabilities, making it possible to study larger, intact samples in unprecedented detail. To overcome limitations like shallow penetration and light diffraction in traditional LSFM setups, advanced beam shaping with devices such as spatial light modulators and digital micromirror arrays has been utilized. These improve the resolution and the extent of tissues that can be imaged. Advances such as Bessel-beam-based LSFM and lattice light sheet microscopy increase the field of view that can be imaged with low background noise, but often require complex and bulky equipment. Addressing these complexities, a new approach builds on silicon nitride photonic integrated circuits to create a structured light sheet in a very compact device. This system incorporates beam-steering based on wavelength control in a limited tuning range and enables the generation of light sheets with optimized characteristics in regard to thickness and diffraction-length limited penetration depth. Simulations include extensive fabrication tolerance analysis that confirm the practicability of the approach, that can be straightforwardly extended to dual wavelength excitation. This compact, chip-based LSFM system could make high-quality imaging more accessible and transform biomedical instrumentation.

## 1. Introduction

Light sheet fluorescence microscopy (LSFM) stands out as a transformative approach for volumetric three-dimensional (3D) imaging in biomedical research. Its applications cover the study of the structural and functional relationships of biological tissues, the mapping of vascular and neural networks, and the diagnosis of disorders such as tumors by imaging of microenvironments with high precision. It supports the imaging needs in investigation fields such as brain machine interfaces, gene therapy, and regenerative medicine [1-5]. When combined with techniques such as calcium imaging and optogenetics, it allows real-time observation of neuronal activity within its anatomical context and the study of cellular dynamics [6-9]. Traditional 3D imaging methods, which reconstruct 3D images from serially imaged thin tissue sections, suffer from distortion, data loss, and limited applicability to smaller samples. In contrast, LSFM overcomes these limitations by offering rapid imaging suitable for the study of dynamic processes and live cells in biological samples [10-12].

In LSFM, a light-sheet illuminates the sample in a plane perpendicular to the imaging axis and excites fluorophores. It combines fast 3D imaging, achieved by scanning the light sheet through the sample and recording the image with a large field of view (FOV), with a resolution ranging from cellular to subcellular levels. By illuminating only the focal plane of the imaging system, LSFM minimizes out-of-focus excitation, effectively reducing phototoxicity, which is critical for live-cell imaging [13-15]. Unlike confocal microscopy, which uses slower, point-by-point illumination and is prone to photobleaching [16, 17], or two-photon (2P) microscopy, which provides deeper penetration but also results in slower acquisition speeds [18, 19], LSFM illuminates entire planes simultaneously, enabling faster imaging with reduced sample damage due to photobleaching.

To overcome degraded image clarity and resolution caused by light scattering and absorption in dense biological samples, tissue-clearing techniques have been developed, that significantly enhance the capabilities of LSFM in imaging large samples [20-23].

These methods render tissues optically transparent by effectively removing lipids while preserving the structural integrity of the sample. This process reduces light scattering and allows for imaging at depths of several millimeters, making image clarity less dependent on the optical properties of the initial sample.

Therefore, the resolution and size of the acquired image are primarily limited by the optical systems used for illumination and imaging. For the imaging system, a high numerical aperture (NA) with a long working distance is required and cleared tissues are typically imaged with an immersion objective. For the light sheet, extent, thickness, and the reduction of side lobes and stray light outside of the imaging plane are critical parameters.

Several approaches have been developed to generate planar light sheets for LSFM, each offering distinct advantages and trade-offs. Traditional methods, such as using cylindrical lenses, Airy beams, or scanning Gaussian beams with Galvo mirrors are relatively straightforward and facilitate dynamic beam positioning, but are also limited by a more shallow penetration depth due to rapid beam divergence and increased susceptibility to photobleaching [24-26]. In contrast, advanced beam shaping optics, such as digital micromirror devices (DMDs) [27] and liquid-crystal spatial light modulators (SLMs) [28, 29] offer dynamic wavefront shaping and precise control over illumination patterns, which allows enhancing the resolution and reducing scattering in heterogeneous tissues. However, they also result in reduced power translation to the generated light sheet, as a consequence of the complex free-space beam forming setups used in conjunction with these devices [30, 31].

To overcome the limited penetration depth resulting from the diffractive nature of Gaussian beams, Bessel beams have been introduced in LSFM light sheet generation [32, 33]. This technique utilizes the unique non-diffracting and self-healing properties of Bessel beams, which are shared across a range of applications, including sensing [34], optical trapping [35], optical coherence tomography [36], and Raman spectroscopy [37]. Using Bessel beams in LSFM microscopy enables deeper tissue penetration and uniform illumination, making it especially effective for imaging thick samples. An ideal, truly non-diffracting Bessel beam has a cross-section with an infinite extent. However, in practical applications, generated Bessel beams are truncated, resulting in an enhanced but finite diffraction length. Furthermore, the side lobes surrounding the central beam, arising from the radial electric field distribution following a Bessel function, lead to a significant portion of the optical power being distributed off-center, compromising the signal-to-noise ratio (SNR). Lattice light sheet microscopy has addressed background suppression by structuring of the beam into a lattice pattern, delivering superior resolution along the axis of the imaging system at the price of a more complex optical setup involving SLMs or diffraction-based beam patterning and sophisticated alignments [38].

In this work, we present a structured LSFM sheet generation concept based on an optical phased array (OPA) [39] implemented in a silicon nitride (SiN) photonic integrated circuit (PIC) platform [40] adapted to the processing of visible wavelengths for biophotonics applications [41, 42]. This approach enables the generation of a light sheet in a very compact and robust setup with enhanced control over the generated illumination patterns. We begin by introducing a high-level schematic of the light sheet generation setup in the next section, which is followed by a mathematical framework describing the forming and steering of the beam in the plane of the sheet through wavelength tuning. Section 3 reports the results of raytracing simulations applied to the PIC emission, that focus primarily on the impact of the grating emitter (GE) distribution on the PIC surface on key light sheet characteristics such as thickness, width, non-diffracting length, and the suppression of parasitic higher-order beams. Emitters are first modeled as each emitting an ideal Gaussian beam propagating along the surface normal of the PIC. Section 4 dives into the design of a concrete PIC and addresses difficulties arising from using practical GEs with an off-angle emission. To ascertain the practicability of this approach, a detailed process bias tolerance analysis is presented in Section 5, followed by an analysis of the required waveguide coherence length in Section 6. Finally, we generalize the PIC architecture to one capable of generating beams at multiple wavelengths in Section 7.

## 2. PIC based LSFM

Figure 1 shows a general schematic of an LSFM setup with PIC-based illumination and a generic imaging system. The PIC is combined with an illumination lens to generate a structured light sheet that illuminates a cleared sample. The optical imaging system has an optical axis that is perpendicular to the light sheet and uses an objective lens whose focal point is aligned with it. It captures images of the cleared sample at different cross-sections as the light sheet and the focal point are moved through it. Additionally, the imaging system includes an optical filter that suppresses the illumination wavelength while allowing the fluorescent emission to be captured by a charge-coupled device (CCD). For cleared brain tissues, we target the use of the fluorochrome TdTomato, which absorbs light most effectively at 555 nm and emits at 580 nm. Since wavelength tuning is later used for lateral steering of the generated light sheet ($x$-axis), we select $525 \pm 5$ nm as the wavelength range for excitation, which is far enough from the emission wavelength to allow for good filtering of the pump, remains inside the excitation spectrum with an excitation efficiency better than 60% of its peak, and is a wavelength range in reach of commercial tunable lasers. In Fig. 1 and in the following, the $y$- and $z$-axes refer to the (vertical) optical axis of the imaging system and to the (longitudinal) direction of propagation of the light sheet, respectively.

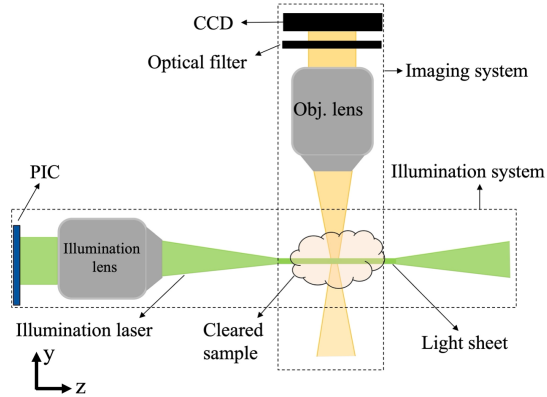

Fig. 1. General schematic of a PIC-based LSFM setup.

Figure 2(a) shows the scaled down schematic of a SiN PIC generating a structured light sheet for LSFM. It operates at a central wavelength of 525 nm and implements scanning of the beam in the $x$-direction by tuning of the wavelength. It is positioned in the rear focal plane of a high NA lens, see Fig. 2(b), that operates a Fourier transform on the generated field. The PIC includes GEs arranged along the circumference of a circle with radius $R$ and a light distribution network formed by cascaded 1-by-2 multi-mode interferometers (MMIs), that is balanced on each side of the GE-array up to the last MMI of the network. The last waveguide segments connecting the last layer of MMIs with the GEs are of variable length, with a length offset equal (or more generally proportional) to the $x$-coordinate of the GE, which, as we shall see, enables the wavelength dependent steering of the beam, or more precisely, of its envelope. A phase difference can be applied to the two sides of the array via thermal phase tuners. They determine the position of optical lobes within the beam's envelope and can be rapidly dithered to homogenize the beam or sequentially switched to implement structured illumination based optical sectioning [38].

The GEs are equidistant in the $y$-direction with adjacent devices offset by $\Delta y = d$, as labeled in Fig. 2(a), leading to a discretization of the targeted annular field emission [43]. This greatly facilitates the balancing of the distribution network, as compared to maintaining a constant azimuthal angle increment along the circular distributions. In conventional OPAs, discretization of the emitted field profile leads to parasitic beams in the form of higher diffraction orders propagating along different directions in a regular $k$-space array. Here, due to the Fourier transform operated by the illumination lens, an excessively coarse discretization associated to a large *d* leads to parasitic light sheets translated in real space to form a periodic array along the *y*-direction, as analytically described in the following and numerically investigated in Section 3.

Figure 2(b) schematically shows the positioning of the PIC, whose emitting surface coincides with the back focal plane of a plano-convex lens used to transform its emission. It also illustrates the expected interferometric pattern in the *xz*-plane, representing the extent of the light sheet as observed from the top-view perspective of the imaging objective, as well as the extent of the light sheet in the *yz*-plane, from which its thickness can be extracted.

As we shall see, the lateral extent of the light sheet along the *x*-direction and its thickness along the *y*-direction can be traded-off against each other by setting the angular coverage *θ* of the GEs on the two sides of the array. As represented in Fig. 2(a), the circle is not fully filled, which would instead lead to the generation of an isotropic Bessel beam. The finite width of the beam can be complemented by rapid wavelength-based steering along the *x*-direction, emulating a wider, homogeneous light sheet. Furthermore, the radius *R* of the GE distribution sets a trade-off between the non-diffractive beam length, setting the depth over which the sample can be imaged, and its thickness.

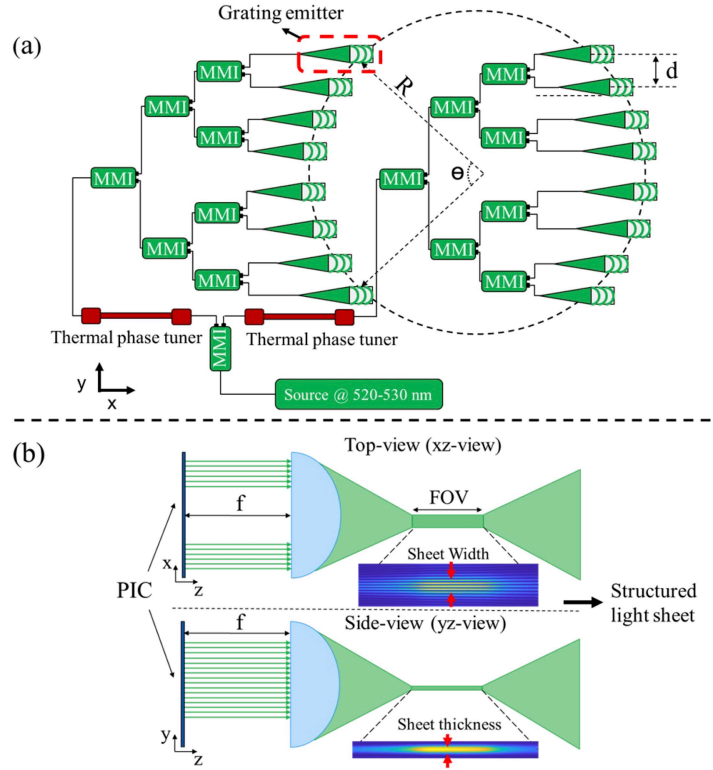

Fig. 2. (a) Simplified architecture of a PIC generating a structured light sheet using GEs arranged in a circular distribution. (b) Top- (*xz*) and side- (*yz*) views of the beam former comprising a PIC located at the rear focal plane of a plano-convex lens, as well as representative light intensity profiles.

Since the PIC is located in the rear focal plane of the lens, locations on the PIC from which light is emitted are mapped to specific *k*-vectors after the lens. Specifically, coordinates (*x*, *y*) are converted into ($k_x$, $k_y$) with $k_\mu = -\sin(\text{atan}(\mu/f_d)) \cdot nk_0 \simeq -(\mu/f_d) \cdot nk_0$, $\mu = x, y$, $k_0 = 2\pi/\lambda$, $f_d$ the focal length of the lens, $\lambda$ the wavelength of the light in vacuum, and *n* the refractive index of the medium after the lens in case of an immersion objective (in the following numerical studies, a non-immersion lens is used for beam forming and *n* = 1). The *k*-vector component along the optical axis of the lens, $k_z$, can then be obtained by $k_z = \sqrt{n^2 k_0^2 - k_x^2 - k_y^2}$. It follows from this equation that for a

GE arrangement on a circle centered on the optical axis of the lens, $k_z$ is identical for all the field components, assuming the dimensions of the GEs to be very small compared to the focal length $f_d$. The total field after the lens can then be obtained as a summation over the index $m$ of the GEs as

$$E(r,t) = \sum_m A_m e^{i(k_{x,m}x+k_{y,m}y+k_z z-\omega t)} = e^{i(k_z z-\omega t)} \sum_m A_m e^{i(k_{x,m}x+k_{y,m}y)} \quad (1)$$

where $A_m$ is the amplitude emitted by the GE of index $m$ and $\omega$ is the angular frequency of the light.

Factoring out the phase dependence in the $z$-direction shows that the beam shape remains invariant as it propagates forward, i.e., it is non-diffracting akin to a Bessel beam. In fact, completely filling the circle serving as a locus for the GEs with a continuous distribution of emitters sharing a common phase $\varphi_m = 0$ would result in the generation of a lowest-order Bessel beam as commonly used in LSFM.

In practice, the size of the GEs along the radius of the circle creates a spread in the generated $k_z$ component distribution and limits the length over which the beam stays collimated. Applying a first-order Taylor expansion to the expression of $k_z$, it can be approximated as

$$\Delta k_z \simeq (R \cdot \Delta R/f_d^2) \cdot nk_0, \quad (2)$$

where $\Delta R$ is the extent of the GEs in the radial direction. The propagation length over which the beam remains collimated, $w_z$, scales as the inverse of this quantity. If $\Delta R$ represents the $1/e^2$ intensity diameter of the Gaussian beam emitted by a GE (also referred to as its mode field diameter in the following), the corresponding full width at half maximum (FWHM) can be predicted as $4\sqrt{2ln(2)}/\Delta k_z$. An alternate derivation of the same formula consists in first considering the size of the beams generated by each GE at the lens and deriving the length over which they overlap at their crossing point, after collimation, around the front focal point.

A small $\Delta R$ is thus beneficial to extend the length of the sheet. However, the fraction of power contained in the center of the beam also scales with $\Delta R$ [44], so that this comes at the expense of the power contained inside the sheet. This is a consequence of the power contained in Bessel-like beams not decaying at large radii if integrated over the corresponding circumference, so that truncation of the beam is required to maintain a finite portion of the power in the central lobe. As also numerically investigated in Section 3, the peak intensity of the generated beam, on the optical axis of the lens, is predicted by $N \cdot P/S$, with $N$ the number of GEs, $P$ the total power emitted from the surface of the PIC and $S = (\pi/2) \cdot (MFD/2)^2$ the effective area of the collimated Gaussian beams after the lens (defined here as the ratio of the power transported by a single Gaussian beam and its peak intensity). There is thus a trade-off between $S$, that results in a large non-diffractive propagation length and a wide steering range (see below) and the power translation efficiency of the beam forming scheme.

The consequences of discretizing the $k$-vector distribution because of the periodic GE arrangement along $y$, with a pitch $d$, can now also be quantified. This results in the $k$-vectors to be sampled on a grid with a period $\Delta k_y = (d/f_d)nk_0$ and thus in copies of the main beam offset in real space by $\Delta y = 2\pi/\Delta k_y = f_d \lambda/dn$. Choosing $d$ to be sufficiently small allows moving these outside the area $S$ illuminated by the beams generated by the GEs, i.e., outside the form factor of the OPA, and suppresses their generation.

Bessel beams, while being non-diffractive and self-healing and thus providing an FOV extending deep into the biological tissue, also suffer from side lobes that smear out the point spread function (PSF) of the imaging system. This has been improved by lattice light sheet microscopy [38], in which a structured light sheet is generated in the form of an optical lattice restricted to a thin sheet extending orthogonally to the axial direction of the imaging objective. However, the positions of the light sources required to generate a conventional lattice light sheet deviate significantly from being on a single circle, so that this also significantly restricts the collimation length of the sheet. For this reason, we pursue a different approach here, that consists in restricting the positions of the GEs to be on a single circle, but also restricting the distribution to two symmetric arcs of circles parameterized by an angular extent $\theta$ [45], as illustrated in Fig. 2(a). This restores the collimation length $w_z$ in the $z$-direction, while creating a structured light sheet of finite extent $w_x$ in the $x$-direction, limited by the generated $k_x$-components now forming a continuum, beyond which the light sheet rapidly broadens. The extent of the sheet in the $x$-direction and its thickness in the $y$-direction can be traded off in an application specific manner by choosing $\theta$, as discussed quantitatively in Section 3. The larger $\theta$, the wider the range of generated $k$-vector components in the $y$-direction ($k_y$) and the thinner the sheet. However, a larger $\theta$ also leads to the range of $k$-vector components in the $x$-direction, $k_x$, to become larger for each side of the OPA, and thus to a more rapid smearing of the beam shape along $x$, i.e., in a reduction of $w_x$.

An alternate path to reducing the thickness of the sheet and thus to improve the axial resolution consists in expanding the radius $R$ of the GE distribution, which increases the effective NA of the beam-forming optics. However, a larger $R$ also leads to a reduction of the collimation length, as predicted by Eq. (2). Moreover, $R$ is limited by the entrance pupil of the illumination lens and by possible aberrations resulting from rays that are originating far from the lens's optical axis. This constraint becomes even more restrictive when the GEs are emitting at an angle relative to the surface normal of the PIC, as discussed in Section 4, as is the case for practical grating couplers ($\alpha \neq 0$) [46]. Furthermore, the finite waveguide coherence length on the PIC limits the waveguide length in the distribution network and thus the maximum $R$ that can be practically implemented, which is analyzed in Section 6.

To maintain a high axial resolution, imaging should be restricted along $x$ to the range in which the light sheet remains thin. To image beyond that range, the beam first needs to be translated, which can be simply done by wavelength scanning as described below. Given the ease with which the beam can be scanned with our method and the possibility to further increase the imaging range along $x$ by displacing the objectives or the sample, this appears preferable over a limited collimation length, which fundamentally limits the penetration depth of the illumination system and thus the size of the biological sample that can be imaged.

The lateral steering of the beam takes the form of a true translation after the lens (as opposed to an angle of emission change as in a conventional OPA). The translation is achieved by tuning the excitation wavelength of the laser within the absorption spectrum of the fluorophores with which the biological sample is stained. The optical path length between the edge coupler (EC) serving as an input port to the PIC and any of the GEs on a given side of the array is the same up to the output ports of the last layer of MMIs, so that all of these are supplied with light with an equal intensity and phase (see Sections 5 and 6 for an analysis in regard to process biases and to the coherent length of the waveguides). The last segments connecting the MMIs to the GEs have a path length difference $\Delta l_m$ proportional to the $x$-coordinate of the GEs, simply taken here as $\Delta l_m = x_m$, since this leads to the simplest routing. The phase $\varphi_m$ of the light emitted by the GE of index $m$ is thus subjected to a phase offset $\Delta \varphi_m$ proportional to the wavenumber $\beta$ of the waveguide and to the $x$-coordinate $x_m$ of the GE and is given by

$$\Delta \varphi_m = \frac{2\pi n_{eff}}{\lambda} x_m \quad (3)$$

with $n_{eff}$ the effective index of the waveguide.

At the center wavelength of the tuning range, $\lambda_0$, we require $\Delta \varphi_m$ to be uniformly set to zero, modulo $2\pi$, so that at this wavelength the beam is centered relative to the optical axis of the beam forming lens. This is achieved by applying a small displacement $dx_m$ to the position of each GE, and thus a small change to the path length offset $\Delta l_m$, that is below a half wavelength in the waveguide $\lambda_0/2n_{eff}$ and is given by

$$dx_m = -\frac{\left[\text{mod}\left(\frac{2\pi n_{eff}}{\lambda_0}x_m + \pi, 2\pi\right) - \pi\right]}{2\pi}\frac{\lambda_0}{n_{eff}} \tag{4}$$

After this correction, the phase offset $\Delta\varphi_m$ becomes

$$\Delta\varphi_m = \frac{2\pi n_{eff}}{\lambda}(x_m + dx_m) \simeq -\frac{2\pi n_g \Delta\lambda}{\lambda_0^2}x_m \tag{5}$$

where $\Delta\lambda = \lambda - \lambda_0$ is the detuning of the excitation wavelength from the center wavelength and $n_g$ is the group index of the waveguide.

The phase offset $\Delta\varphi_m$ is thus proportional to $x_m$ and thus to $k_{x,m}$, since $k_{x,m} \simeq -(x_m/f_d) \cdot nk_0$. This means that $-\partial\Delta\varphi/\partial k_x$ is a constant $\Delta x$ and the envelope of the beam is simply displaced by that amount along the $x$-direction (a simple property of Fourier transforms and envelope functions, analogous to a group delay $\tau = \partial\varphi/\partial\omega$ introduced in the time domain with the phasor convention used in Eq. (1)). This can be further simplified into

$$\Delta x = -\frac{2\pi n_g \Delta\lambda}{\lambda_0^2}\frac{f_d}{nk_0} = -\frac{\Delta\lambda}{\lambda_0}\frac{n_g}{n}f_d \tag{6}$$

Up to here, we have assumed that the GEs emit beams along the surface normal of the PIC with an angle $\alpha = 0$. However, this is known to result in grating couplers with very poor coupling efficiency and is thus not practical. In practice, the GEs will be designed to emit with a slight angle. This results in the two groups of beams generated after the lens by the left and right side of the OPA to cross at a point that is away from the lens's optical axis (see Fig. 3 in the next section). The targeted interference pattern can only be generated where these two beams cross and interfere, akin to the form factor of a simple grating. The location of the interference pattern, as described above (the structure factor) needs thus to be recentered to the crossing point of these two beams. This can be achieved by updating Eq. (4) with this constraint. We first establish the displacement $\Delta x_0$ by which we need to shift the interference pattern at the center wavelength $\lambda_0$ by determining the crossing point of the beams from geometrical optics. From this, we extract the required (non-zero) phase offset $\Delta\varphi_m(\lambda_0)$ at the center wavelength from $\Delta\varphi_m(\lambda_0) = -\Delta x_0 \cdot k_{x,m} = \Delta x_0 \cdot (x_m/f_d) \cdot nk_0$. Finaly, we update Eq. (4) as

$$dx_m = -\frac{\left[\text{mod}\left(\frac{2\pi n_{eff}}{\lambda_0}x_m - \Delta\varphi_m(\lambda_0) + \pi, 2\pi\right) - \pi\right]}{2\pi}\frac{\lambda_0}{n_{eff}} \tag{7}$$

It should be noted that the steering range in which the beam can be translated is not only limited by Eq. (6) and the tuning range of the laser, but also by the targeted beam position (the interference point or structure factor) moving in and out of the area $S$ around the beam crossing point shown in Fig. 3 (the form factor limited by the finite emission cone of the GEs, or equivalently their non-zero extent $\Delta R$). This can be improved by exploiting the wavelength dependence of the GE emission angle $\alpha$. If the interference pattern is shifted to the right by a positive $\Delta x$, it is advantageous to also move the beam crossing point in the same direction by increasing $\alpha$. This is exactly what happens in the configuration shown in Fig. 2(a) if the GEs are implemented as forward scattering grating couplers (i.e., emitting light with a positive $\alpha$, i.e., with a positive $k_x$), which is the configuration used in the following. This can be further optimized by designing the dispersion inside the grating couplers such that the beam crossing point moves exactly as $\Delta x$ as the wavelength is scanned. Such dispersion engineering has for example been utilized to stabilize the beam emission angle and obtain wideband fiber coupling [47, 48]. Here, it could be used to obtain the targeted dependence.

## 3. Raytracing Simulations

In this section, we first perform a set of simplified raytracing simulations in which we assume the GEs to be vertically emitting. The effects of the radius $R$, the angular coverage $\theta$ and the GE-to-GE spacing along the y-direction, $d$, are systematically investigated. Moreover, we validate by simulation the placement of the GEs according to Eq. (4) and the wavelength dependent beam scanning range.

Figure 3 shows the top-view ($xz$-plane) schematic of the raytracing model created in Zemax OpticStudio. The model includes 512 GEs, each modeled as emitting a Gaussian beam with a $1/e^2$ mode field diameter (MFD) of 5.5 μm at its waist, which corresponds to the closest Gaussian emission profile approximation of the concrete GE design presented in the next section. The GEs are positioned on the circumference of a circle of radius $R$ located on the back focal plane of a convex lens (Thorlabs A240) and centered on the lens's optical axis, which has a rear working distance (WD) of 5.92 mm and an effective focal length $f_d = 8$ mm. For the calculation of the emission phases $\varphi_m$, we consider a waveguide effective index of $n_{eff} = 1.61$ at 525 nm and a group index of $n_g = 1.88$, obtained from the optical mode simulation of a double stripe SiN waveguide platform [40], see Section 4 for exact layer thicknesses.

The beams emitted from the GEs are collimated by the lens, after which they intersect in its forward focal plane. This creates the targeted interference pattern in the $xy$-plane, as shown in the inset Fig. 3(b). Inside the simulation, around the region where the beams cross, the generated field profile is recorded by 200 detectors oriented in the $xy$-plane and displaced along $z$ to record different field cross-sections along the sheet's axis of propagation. Each detector has a resolution of 200 by 200 points and a size varying from 20 by 20 μm to 150 by 150 μm depending on the simulation, that can be identified in the 2D color plots of the intensity profiles, that show the entire range of the recorded data. The spacing along $z$ is 10 μm in most simulations to span a length of 2 mm along the $z$-direction. The field dependency along $z$ is recovered by concatenating the data from the detectors. Simulations are run in non-sequential mode so that interference effects can be modeled and 1000 rays are launched for each GE, from which the Huygens point-spread-function (PSF) is calculated in the detector plane.

Since the lens is not an ideal thin lens, it features some level of aberration, apparent in the beams emitted by single GEs not being perfectly collimated after the lens if the PIC is placed in its rear focal plane. This, however, could be corrected by reducing the distance between PIC and lens relative to the nominal working distance by an amount dependent on $R$, after which near perfect collimation could be recovered. This is exemplified by Fig. 3 that shows the raytracing setup for $R = 3$ mm. Optimum results were obtained with PIC-to-lens distances of 5.51 mm and 3.96 mm, for $R = 1$ mm and $R = 3$ mm, respectively.

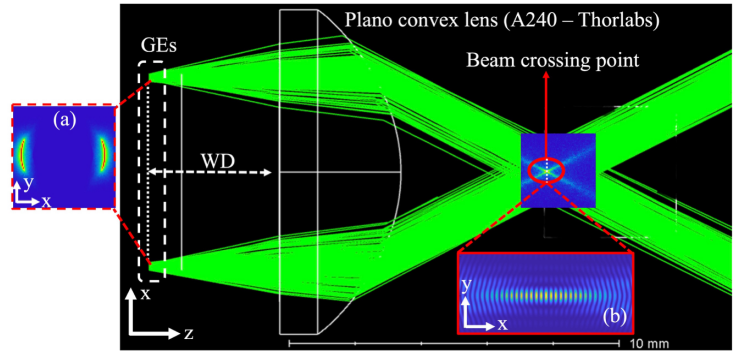

Fig. 3. $xz$-view of the raytracing simulation. GEs are located at the rear focal plane of the convex lens and the beams emitted from the PIC cross each other at a location on the forward focal plane. Insets (a) and (b) show the intensity profiles in the $xy$-plane at the surface of the PIC and at the beam crossing point, respectively.

To benchmark our beam forming concept, we first simulate non-diffracting Bessel beams generated by circular GE distributions with full angular coverage distributed over a circle of radius $R = 3$ mm, shown in Fig. 4, panel A. Figure 4(a) shows the generated interference pattern in the *xy*-plane, that takes the expected form of a Bessel beam. Figures 4(b)-4(d) display the intensity distributions along cuts across the *x*-, *y*- and *z*-axes. The full width at half maximum (FWHM) of the central lobe along the *x*- and *y*-axes, $w_{x,y}$, that is representative of the thickness of the sheet obtained from scanning the beam along *x*, is determined to be 0.4 µm. The collimation length along *z*, quantified by the FWHM along the beams' propagation direction, $w_z$, is 665 µm.

Panel B shows the simulated intensity profiles for our beam forming concept with GEs distributed over the same circle, but restricted to two arcs with an angular coverage of $\theta = 120°$. This extends the lateral extent $w_x$ to 2.5 µm without a significant increase of $w_y$, that is now 0.6 µm. This is the consequence of the range of *x*-coordinates, and the associated spread of generated $k_x$ vector components, being significantly reduced on both sides of the array, while the range of *y*-coordinates still fills most of the original extent. At the same time, the side-lobes along *y* are significantly suppressed above and below the sheet's central axis at $x = 0$, which is an important advantage. Spreading the intensity over a small sheet of a few µm extent along *x* also offers the advantage to reduce the peak instantaneous intensity while maintaining the same average intensity as the beam is scanned. This reduces photobleaching while maintaining signal-to-noise during the imaging process if the pixel acquisition at the imager is synchronized with the scanning of the beam. Here, $w_x$ is extracted from the envelope function shown by the dashed blue line, that is obtained by averaging the intensity obtained from the phase shifters in Fig. 2(a) applying (i) the same phase and (ii) a π phase offset relative to each other, i.e., by emulating rapid dithering of these phase shifters for homogenization of the beam envelope.

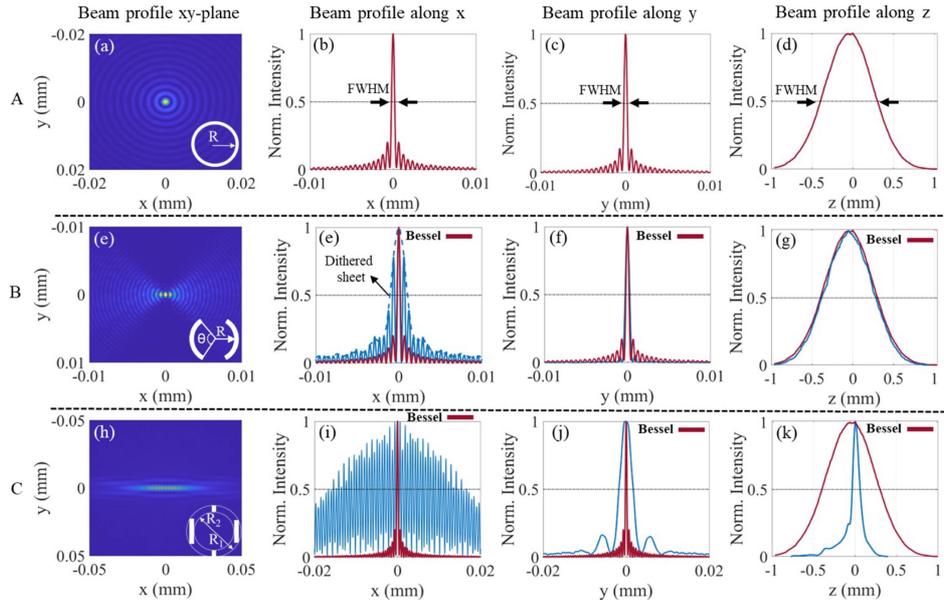

Fig. 4. Intensity distribution of beams generated with discrete GEs arranged on a circle. (A) Bessel beam ($R = 3$ mm), (B) structured light sheet ($R = 3$ mm, $\theta = 120°$), and (C) triangular lattice ($R_1 = 3$ mm, $R_2 = 2.6$ mm). (a), (e), and (h) show the intensity distribution in the *xy*-plane with the GE distribution shown in the inset. (b), (e), and (i) show the intensity along the *x*-axis, (c), (f), and (j) along the *y*-axis, (d), (g), and (k) along the *z*-axis. In B and C, the *x*-, *y*-, *z*-axis intensity profiles (in blue) are overlaid with those of the Bessel beam (in red). The dashed blue line in (e) shows the intensity profile obtained by dithering the relative phase between the two sides of the OPA.

In panel C, we present the beam formed by GEs distributed along four straight lines, as represented in the inset of Fig. 4(h), generating a triangular lattice light sheet, corresponding to what is more commonly done in lattice light sheet microscopy. These lines are extended between two concentric circles with radii $R_1 = 3$ mm and $R_2 = 2.6$ mm, which determines the range of $k_y$-components for each distribution and thus the thickness $w_y$ of the sheet. The GEs on the right- and leftmost distributions are at $x = \pm 2.8$ mm, the midpoint between the two circles, and exhibit an expansion along the y-axis of 2.1 mm.

The GE distributions at the north and south are centered at $x = 0$ mm and each extend along 0.4 mm. This configuration results in a much wider sheet with $w_x = 32$ µm. However, the beam thickness, $w_y$, increases to 3.5 µm, while the beam's penetration depth, $w_z$, reduces to 79 µm. The trade-off between these two quantities depends on the difference between $R_1$ and $R_2$ and does not play a role for the concept illustrated in panel B, since there the GEs remain on a single circle.

In the following, we focus on configurations in which the GEs are located on two continuous arcs located on the two sides of the circle, akin to the configurations shown in Figs. 2(a) and 4(B). Figure 5 compares two sets of simulations in which the radius is changed between 1 mm (upper panels) and 3 mm (middle panels), while the angular coverage of $\theta = 120°$ is maintained. In panels 5(a) and 5(d), the illumination pattern is shown in the $xy$-plane at the forward focal plane of the lens. The width of the sheet, $w_x$, as quantified by the FWHM of the envelope function obtained when applying dithering, ranges from 7 µm for $R = 1$ µm to 2.5 µm for $R = 3$ mm. Panels 5(b) and 5(e) display the intensity distribution in the $xz$-plane (top-view), illustrating a concomitant decrease in the penetration depth, $w_z$, from 5.3 mm to 665 µm. However, at the same time the thickness of the sheet in the axial direction of the imaging objective also decreases from 2 µm (Fig. 5(c)) to 0.6 µm (Fig. 5(f)).

Figures 5(g)-5(i) illustrate the overlay of the intensity distributions resulting from the two radii along the $x$-, $y$-, and $z$-axes, highlighting the shrinkage in the beam's width, thickness, and length.

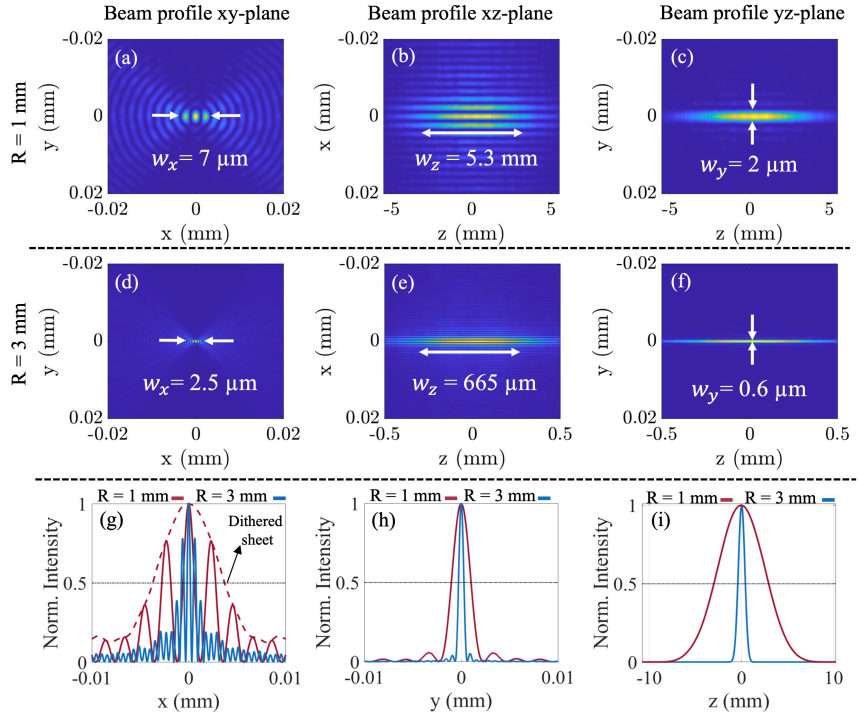

Fig. 5: Intensity distributions of the generated light sheets in (a), (d) the $xy$-plane (cross-section), (b), (e) the $xz$-plane (top-view), and (c), (f) the $yz$-plane (side-view) for a radius $R = 1$ mm (upper panels) and $R = 3$ mm (middle panels). Overlaid intensity profiles for $R = 1$ mm (red) and $R = 3$ mm (blue) across the (g) $x$-, (h) $y$-, and (i) $z$-axes. The dashed red line in (g) shows an exemplary envelope obtained from dithering the phase shifters for the $R = 1$ mm configuration. $\theta$ is maintained at 120° in both.

To study the impact of angular coverage on beam characteristics, we present two simulations in Fig. 6 with varying angular coverage. The top and middle panels of Fig. 6 display structured light sheets with GE distributions of common radius $R = 3$ mm but differing angular coverage, with $\theta = 60°$ and $\theta = 120°$, respectively. Again, panels (g)-(i) represent the overlay along the three axes. As expected, increasing the angular coverage results in a more confined light distribution along both the $x$- and $y$-axes, while the collimation length along the z-axis remains unchanged due to the common radius of the two distributions. Thus, varying $\theta$ provides a trade-off between the width and the thickness of the sheet. By increasing $\theta$ from 60° to 120°, the width of the light sheet is reduced from 7.4 μm to 2.5 μm, but its thickness is also improved from 1 μm to 0.6 μm.

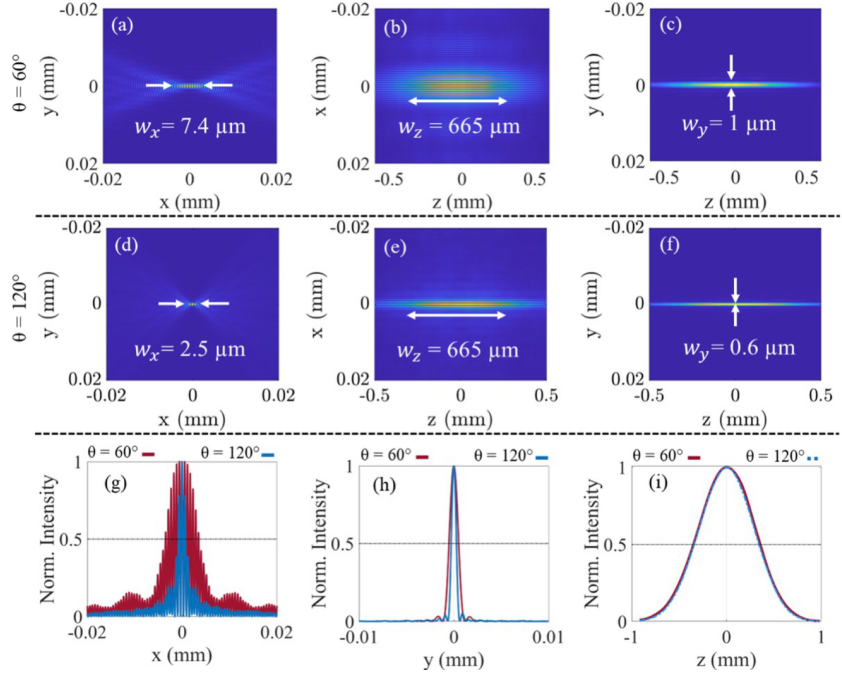

Fig. 6: Intensity distributions of the generated light sheets in (a), (d) the xy-plane (cross-section), (b), (e) the xz-plane (top-view), and (c), (f) the yz-plane (side-view) for $\theta$ = 60° (upper panels) and $\theta$ = 120° (middle panels). Overlaid intensity profiles for $\theta$ = 60° (red) and $\theta$ = 120° (blue) across the (g) x-, (h) y-, and (i) z-axes. R is maintained at 3 mm in both configurations.

Figure 7 summarizes the data obtained when both the radius $R$ and the angular coverage $\theta$ are varied. Panels 7(a)–7(c) show the width ($w_x$), thickness ($w_y$) and collimation length ($w_z$) of the light sheet as a function of the angular coverage $\theta$ and for different radii $R$. As previously mentioned, increasing the angular coverage and the radius causes the illumination pattern to shrink in both the lateral and vertical axes. Meanwhile, the collimation length ($w_z$) only depends on $R$ and decreases from 5.3 mm to 665 μm as the radius is increased from $R$ = 1 mm to $R$ = 3 mm. It should be noted that $w_x$ differs from $w_y$ even for $\theta$ = 180°, since dithering is applied to extract $w_x$ which breaks the symmetry of the beam even in that case.

Figure 7(d) displays the ratio $w_x/w_y$ as a function of $\theta$ for different $R$ and shows that this ratio, and thus the trade-off between $w_x$ and $w_y$, only depends on $\theta$, while $R$ serves to trade off the absolute values of $w_x$ and $w_y$ with $w_z$. Finally, Fig. 7(e) displays $w_y$ as a function of $w_x$ for different $\theta$ and $R$, which is implicitly varied in the curves. This shows that a smaller $\theta$ is generally conducive to obtain both a large width, $w_x$, and a small thickness, $w_y$. However, for a fixed lens NA that bounds the largest $R$ that can be implemented, decreasing $\theta$ also limits the lowest sheet width that can be obtained, and the thinnest sheets are obtained for the largest $\theta$ at fixed $R$. This is why we primarily focus on 120° here.

The choice of parameters can be further optimized on an application specific basis. For example, to achieve a subcellular axial y-resolution with a sheet thickness of ~1 μm, suitable for imaging intra-cellular structures such as mitochondria or the Golgi apparatus, an angular coverage greater than $\theta$ = 70° and a ring radius of $R$ = 2.5 mm are required. This results in a light sheet with a width ($w_x$) of 6.5 μm and a collimation length ($w_z$) of 1.25 mm. Conversely, for a sheet thickness of 5 μm, which is reasonable for imaging entire neurons or retinal cells, using a smaller ring radius of 1 mm results in a sheet width of 40 µm and a collimation length of 5.3 mm for an angular coverage of $\theta = 50°$.

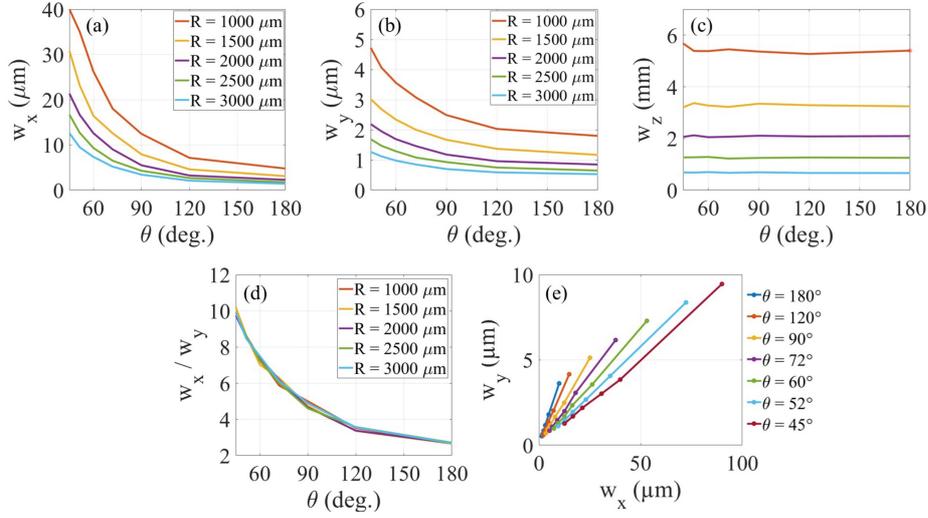

Fig. 7. Light sheet (a) width, $w_x$, (b) thickness, $w_y$, (c) collimation length, $w_z$, and (d) ratio $w_x/w_y$ as a function of angular coverage $\theta$ for different radii $R$. (e) $w_y$ as a function of $w_x$ for different $\theta$. $R$ is implicitly varied in each of the curves in (e).

For further investigations, we choose the configuration with $R = 3$ mm and $\theta = 120°$. In the following, we show that wavelength tuning between 520 nm and 530 nm allows lateral scanning of the beam in a range of ± 98 µm, so that a total FOV of 196 µm by 665 µm is in reach without having to move any parts, which is significant as it is also the typical range that can be captured by a high-NA 100x objective on the imaging side. This steering range is predicted by Eq. 6 and scales with the focal length of the lens, $f_d$, based on which the latter was selected. In addition to verifying this equation, the following raytracing simulations also serve to verify that the power inside the beam is maintained to a sufficient high level during the steering, i.e., that the interference pattern remains in the area $S$ in which the beams from the individual GEs cross and interact.

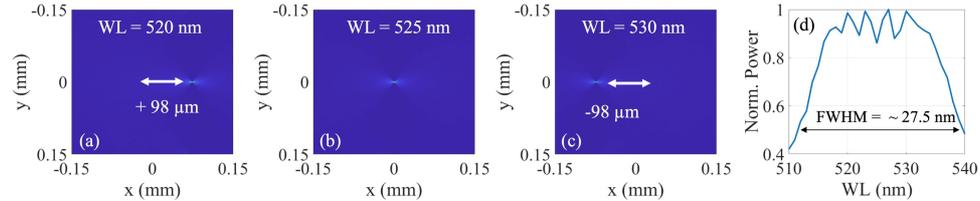

Fig. 8. Beam steering obtained from wavelength tuning for the idealized $\alpha = 0$ case. Intensity in the front focal plane at (a) 520 nm, (b) 525 nm, and (c) 530 nm. (d) Power contained in the central lobe of the beam as the wavelength is tuned, relative to the power when it is at the central position.

Figure 8 presents the simulation results. The wavelength is scanned in a range from 520 to 530 nm. Panel (b) shows the beam at the central wavelength, while panels (a) and (c) display the steered beam at 520 nm and 530 nm, respectively, demonstrating a ±98 µm shift of the beam envelope along the lateral $x$-axis without any other noticeable changes in the beam profile. Panel (d) presents the power contained in the main lobe during beam steering as the wavelength is tuned. It is integrated in a rectangular area covering twice the interference pattern's FWHM in the $x$- and $y$-directions (2×$w_x$ by 2×$w_y$, which is 5 µm by 1.2 µm) and is normalized relative to the power obtained for the central position. It can be observed that the power contained in the central lobe of the interferometric patterns remains close to its maximum over the entire 520 to 530 nm steering range. This is expected since the MFD of the beams emitted by individual GEs, after collimation by the lens, is ~475 µm, which is quite large compared to the induced displacement range. Increasing the utilized excitation wavelength range and steering range further, on the other hand, does not only lead to a reduced fluorophore absorption cross-section, but also in a drop in the power translation efficiency, as the interferometric pattern is steered out of the beam crossing point.

In absolute terms, the power contained in the central lobe at the center wavelength is only 0.2% of the total power transmitted by the lens, which is quite low. This is a common challenge with Bessel-like beams, besides their many advantages. Since their power, integrated over the circumference of a circle centered on the beam's axis, does not decay with radius, only their truncation at large radii makes them physical and the larger the cutoff radius, the smaller the percentage of the power contained in the central lobe. However, truncating the beam too early also reduces the steering range and the divergence free propagation length. As expected, the power translation observed here is in the order of $N \cdot w_x w_y / S$, and can be improved by reducing $S$, which plays the role of the truncation described above and can be achieved with GEs emitting beams with a larger waist and thus a smaller divergence angle. Increasing the number of GEs, $N$, also increases the power translation efficiency and is also explored below, with 1024 providing close to full coverage along the targeted arcs of circle on the PIC:

The last parameter to be examined in this section thus consists in the number $N$ of GEs forming the OPA. It is an essential parameter, as excessively increasing the spacing $d$ between the GEs in the $y$-direction leads to the generation of multiple copies of the beam at offsets along the $y$-axis that are multiples of $\Delta y = f_d \lambda / dn$, as derived in Section 2. If these fall within the area $S$, which is the case at large $d$, they also carry substantial power.

We performed three simulations with a total of 128, 256, and 512 GEs arranged on a circle with $R = 3$ mm and $\theta = 120°$ (the nominal configuration selected above). These result in $d$ to be 82.5 µm, 40.9 µm and 20.4 µm, respectively. The intensity profiles of the generated beams, recorded in the $xy$-plane, are shown in Fig. 9. For the case of 128 GEs, eight side lobes are visible at distances of 40, 80, 120 and 160 µm above and below the main lobe (Fig. 9(a)). For 256 GEs, the spacing between the beams is increased, and only four side lobes are observed (Fig. 9(b)). Finally, for 512 GEs, the default number used so far, the secondary beams are moved further outward (Fig. 9(c)). To ensure secondary beams are spaced far apart and do not carry substantial power, we further increase the number of GEs to 1024 in the following sections. This adjustment shifts the nearest side beams to ±320 µm, at which the intensity of the Gaussian beams generated by individual GEs has already decayed to ~2.6% of its peak, preventing the generation of secondary beams with substantial power. Not surprisingly, this also corresponds to an OPA in which the targeted emission regions are almost fully covered (see Fig. 10 in the next section), as this is the usual condition for the suppression of side beams. The layout requires nine layers of one-by-two splitters on each side of the array and is presented in the next section. Every doubling of $N$ also results in a doubling of the power translation, with 0.4% obtained at $N = 1024$. Further improvement of the power translation requires GEs with a smaller divergence angle or a lens with a shorter focal length, to reduce $S$.

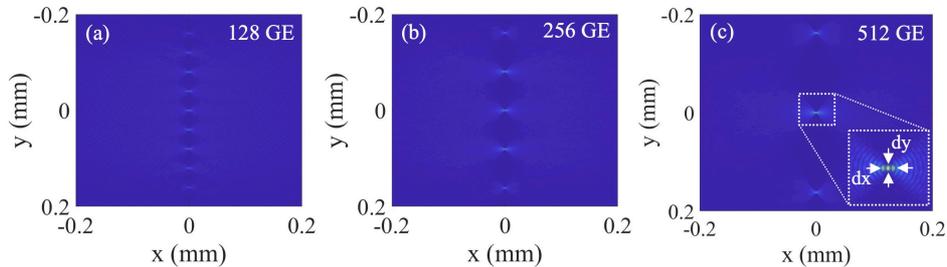

Fig. 9. Emission profiles in the front focal $xy$-plane of the lens, with $R = 3$ mm, $\theta = 120°$ and a total of (a) 128, (b) 256, and (c) 512 GEs. The inset in (c) shows an enlarged view of the beam's central lobe, highlighting the integration area bounded by dx = $2w_x$ and dy = $2w_y$.

## 4. PIC implementation

We evaluate a concrete PIC designed into a customized version of the double stripe waveguide platform provided by Lionix International BV [40]. It features stacked SiN layers, which are near stoichiometric $Si_3N_4$ deposited by low pressure chemical vapor deposition (LPCVD) and have thicknesses of 27 nm (bottom) and 80 nm (top), separated by a 50 nm silicon dioxide ($SiO_2$) interlayer and are also entirely clad with $SiO_2$. We design the chip for operation with transverse electric (TE) modes, that facilitate the design of grating couplers. The architecture requires low loss MMIs and compact GEs to be designed into the platform. MMIs have been designed with less than 0.5 dB of excess insertion loss. Since they are relatively standard devices, we will not describe them further here. The GEs, on the other hand, present some more specific challenges and are discussed in more detail. They need to be compact to increase the non-diffracting path length of the beam ($w_z$) (see Eq. (2)), they need to have a high coupling efficiency towards the top surface of the PIC (a high directionality), and they need to emit light in a direction as close as possible to the surface normal of the chip (small $\alpha$).

Minimizing the emission angle $\alpha$ allows the use of the full NA of the illumination lens. As we shall see, an angled GE emission forces us to reduce the radius $R$ of the GE distribution, which is equivalent to using a smaller aperture, reduces the effective NA of the lens, and thus increases the achievable light sheet thickness and the corresponding axial resolution. The minimum achievable $\alpha$ is constrained by the requirement to avoid the Bragg condition for waveguide-to-waveguide back-reflection (corresponding to perfect vertical emission) over the entire range of utilized wavelengths, 520 nm to 530 nm, and over possible fabrication variations, since reaching the Bragg condition results in a drastic drop in the GE coupling efficiency.

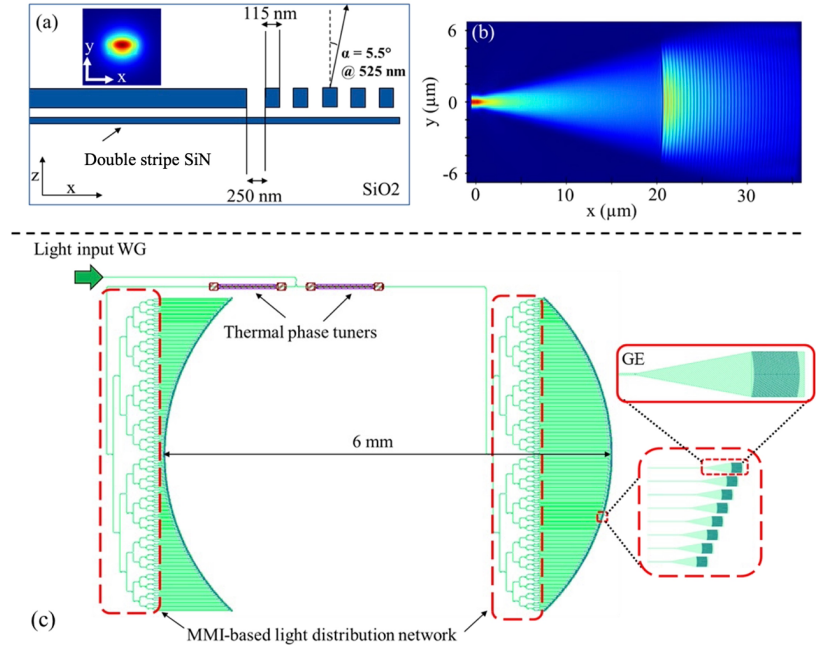

Fig. 10. (a) Schematic side-view of a GE. The inset shows the intensity profile of the TE optical mode for a 400 nm wide double stripe SiN waveguide at 525 nm. (b) Top-view of the simulated light intensity across a focusing GE, recorded in the top $SiO_2$ cladding, 400 nm above the top SiN layer. (c) Layout of a PIC implementing 1024 GEs.

A small $\alpha$ is necessary since the emitted beams undergo a lateral displacement due to their off-axis propagation direction before reaching the lens, that cannot be compensated by displacing the GEs themselves, since they are required to remain symmetrically arranged on a circle centered on the optical axis of the lens. Consequently, GE beams that would have otherwise been adequately collimated by the lens might get clipped at its outer periphery, unless the radius $R$ is further constrained in size (as we shall see, the maximum radius considered so far, 3 mm, is within this constraint). The beam crossing point after the lens is also displaced by the corresponding amount, which needs to be accounted for in the evaluation of Eq. (7).

Figure 10 illustrates the designed GE and shows the layout of a concrete PIC implementing 1024 GEs. Panel 10(a) is a schematic representation of the GE cross-section. Grates are defined by etching 250 nm wide trenches into the top SiN layer, that are backfilled with $SiO_2$. The remaining grates are 115 nm wide, which results in a period of 365 nm. This configuration results in a forward emission angle of $\alpha = 5.5° \pm 0.8°$ in air within the 520 nm to 530 nm wavelength range. The grating is non-apodised, as its grates are already close to allowable minimum feature sizes. The closest Gaussian approximation of the emitted beam has an MFD of 5.5 µm (diameter at $1/e^2$ intensity drop), as also assumed in the raytracing simulations. Figure 10(b) shows simulation results for the entire GE designed as a focusing grating coupler [38] in which the single mode of the 400 nm wide input waveguide is expanded by a taper with a full angle of 22° and in which the first grate is placed 1 µm after the onset of the taper. The GE comprises 35 trenches. According to three-dimensional (3D) finite-difference time-domain (FDTD) simulations, 42% of the input power is scattered towards the top side of the PIC with a Gaussian beam overlap of 75%. Panel 10(c) shows the PIC layout including 1024 GEs on a circular circumference with $R = 3$ mm and $\theta = 120°$, consistent with the main configuration investigated in the previous section.

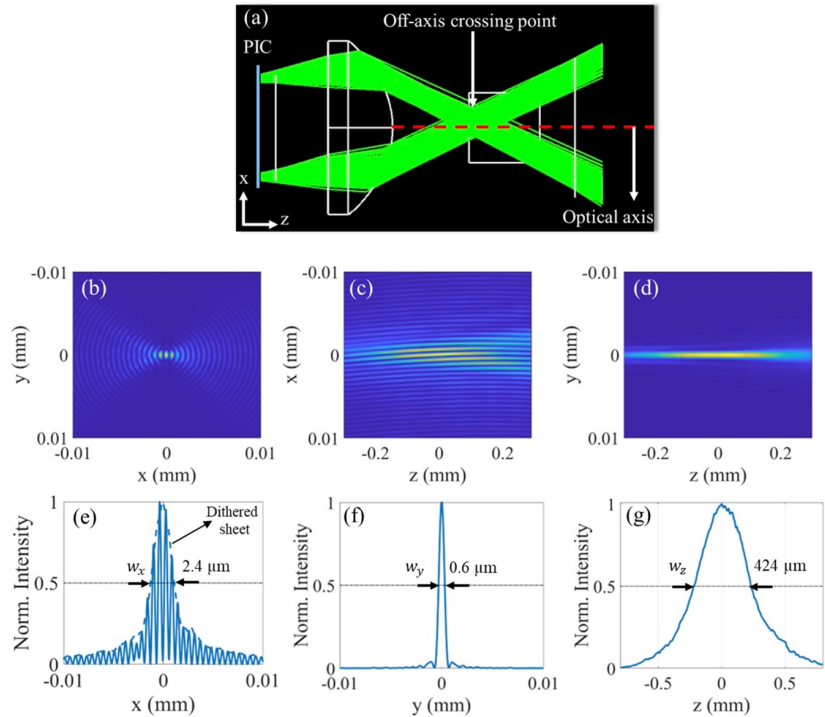

Fig. 11. (a) $xz$-view of the raytracing simulation with non-zero GE emission angle ($\alpha = 5.5°$). The beams emitted from the PIC cross each other at an off-axis point that is displaced in the lateral direction (along the $x$-axis) at the forward focal plane. Intensity distributions for the generated light sheet along (b) the $xy$-plane, (c) the $xz$-plane, and (d) the $yz$-plane for a radius of R = 3 mm and $\theta = 120°$. Intensity profiles of the generated light sheet along (e) the $x$-axis, (f) the $y$-axis, and (g) the $z$-axis. The intensity profile obtained after dithering is overlaid as a dashed line in (e). The intensity profile in (g) is also obtained after dithering, to avoid the slanted orientation of the interference fringes seen in (c) skewing the extracted FWHM. Note that the x-coordinates are recentered such that the central Gaussian beam crossing point is at $x = 0$.

Figure 11(a) displays the corresponding raytracing simulation model featuring the generated rays. It illustrates the off-axis beam crossing point, that is displaced by 385 µm in the x-direction by the finite $\alpha$ at the 525 nm center wavelength. Since the direction of emission is the same for GEs located on both sides of the array, which are oriented in the same way, the symmetry of the beams arriving at the convex lens is broken and they are all shifted in the same direction along the x-axis. As already mentioned, this cannot be compensated by shifting the position of the PIC. The center of the circular GE distribution needs to remain on the optical axis of the lens to satisfy Eq. (1). Instead, the interference pattern is recentered onto the crossing by means of Eq. (7).

Figures 11(b) to 11(g) show the intensity profile of the light sheet that is generated this way. The achieved light sheet thickness $w_y$ is determined to be 0.6 µm, as in the previous section. The other dimensions, $w_x$ = 2.4 µm and $w_z$ = 424 µm are also in line with the previous simulations, with some reduction in the non-diffractive propagation length $w_z$ that results from the aberration of the lens. The beams generated from the two sides of the OPA arrive at the lens at different distances from its optical axis and propagate at slightly different angles after it due to aberration. This is also seen in the interference fringes of the structured light sheet that are now slightly tilted relative to the z-axis, see Fig. 11(c). This further skews the extracted collimation length $w_z$ as an intensity profile along the z-axis would move in and out of such fringes. To counteract this, dithering is first applied to homogenize the light sheet prior to plotting the intensity profile shown in Fig. 11(g).

The GE emission angle $\alpha$ decreases from 6.3° to 4.7° as the wavelength is increased from 520 nm to 530 nm. This leads to a displacement of the beam crossing point towards lower x-coordinates. Since this is also the direction towards which the interference pattern is translated according to Eq. (6), this actually helps maintaining a larger optical power in the sheet as steering is applied, by maintaining the overlap of the generated light sheet with the area $S$ in which the Gaussian beams cross. Figures 12(a)-12(c) show the beam steering obtained by tuning the wavelength within a range of 520 to 530 nm, for the $R$ = 3 mm and $\theta$ = 120° configuration also assumed above. Figure 12(d) illustrates the optical power confined within a window that is twice the beam width and thickness (2 × FWHM), normalized to the power of the central lobe at 525 nm, as previously shown in Section 3 for the $\alpha$ = 0° case. For comparison, The red curve shows a reference simulation in which the emission angle $\alpha$ is artificially kept constant at its average 5.5°, while the blue curve shows the optical power when $\alpha$ varies with wavelength according to the FDTD simulation of the gratings. It is evident that the naturally occuring emission angle variation improves the optical power confined within the sheet, with the FWHM as a function of wavelength increasing to 23 nm from the 13 nm of the fixed-angle case.

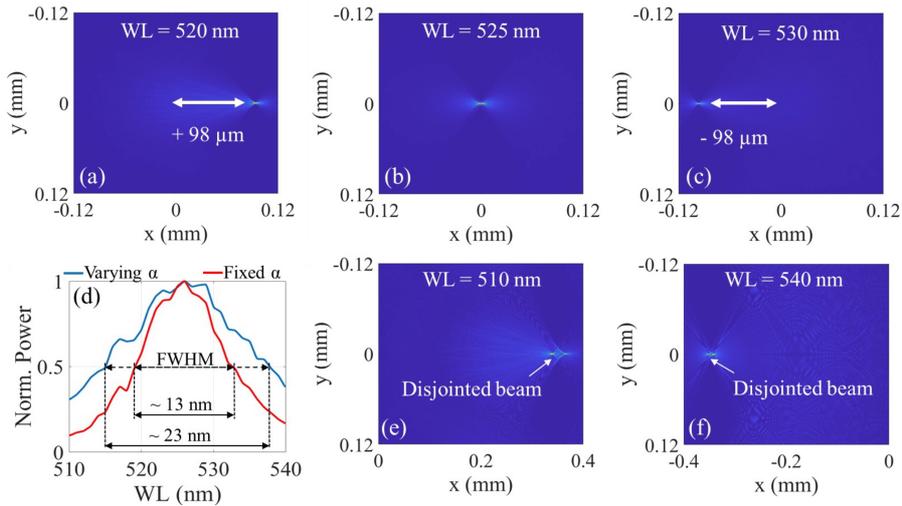

Fig. 12. Beam steering obtained from wavelength tuning with real GEs emitting at a finite $\alpha$. Intensity in the front focal plane at (a) 520 nm, (b) 525 nm, and (c) 530 nm. (d) Power contained in the central lobe of the beam as the wavelength is tuned, relative to the power when it is at the central position. Intensity in the front focal plane at (e) 510 nm and (f) 540 nm showing the disjointed interference patterns at extreme steering due to lens aberration.

However, this still remains below the 27.5 nm observed for the $\alpha = 0°$ simulations in Section 3. This is due to another mechanism limiting the steering range that is apparent at the edges of the extended simulation wavelength range at 510 nm and 540 nm. At these extreme wavelengths, the interference pattern is split into two disjointed beams that are separated by a few µm along $x$ and correspond, each, to one side of the OPA. This is shown in Figs. 12(e) and 12(f) and only occurs for non-zero emission angles $\alpha$, as a consequence of the lens aberation already mentioned above. Rays from the two sides of the array arrive at the lens at different distances from its optical axis. This leads in turn to the rays after the lens being deflected to different angles relative to the optical axis, and thus for some variability in the effective forward focal length of the lens for the two side of the OPA. Since the magnitude of the steering, as prediced by Eq. (6), also depends on $f_d$, this leads to the two interference patterns to separate at large steering distances and to the peak power of the sheet to be reduced. It does not, however, limit steering in the envisioned 520 to 530 nm wavelength range.

## 5. Tolerance to process biases

One of the difficulties faced by a practical implementation of the OPA proposed here is that the correct evaluation of Eq. (7) depends on both the effective index of the interconnect waveguides used to implement the distribution network on the chip and on the exact emission angle of the GEs. Both are subject to fabrication biases that need to be considered in a sensitivity analysis.

Figures 13(a) and 13(b) present the simulated GE emission angle and coupling efficiency as a function of the fabricated trench widths, assuming that the grating periodicity is on target. Due to critical dimensions related to oxide filling in the trench area, we selected a nominal trench width of 250 nm. This choice led us to opt for a remaining grate size of 115 nm and a period of 365 nm to obtain the desired 5.5° emission angle $\alpha$ at 525 nm. While targeting a higher emission angle would have resulted in a marginally larger coupling efficiency, it would also limit the effective NA of the beam forming system, as explained in the previous section. An even smaller $\alpha$, on the other hand, creates the risk of running into a cliff if process variations result in the Bragg condition being reached at the larger wavelength end of the utilized spectrum (530 nm).

Furthermore, we assessed how variations in the thicknesses of both SiN layers and of the $SiO_2$ interlayer affect the emission angle of the GEs, considering a ±10% variation in the thickness of each layer. This resulted in a maximum change of ±0.2° in the GE emission angle, which is a variation in the same order as the ±0.23° already induced by varying the trench width by ±10 nm and has to be compounded with it.

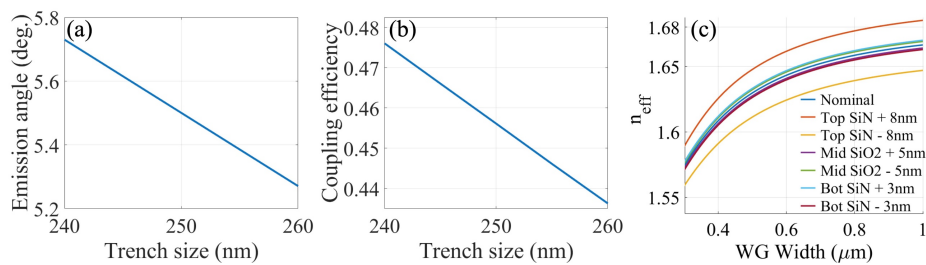

Fig. 13. (a) GE emission angle and (b) coupling efficiency as a function of the trench width at the targeted grating periodicity and 525 nm center wavelength. (c) Effective index of the TE ground mode as a function of waveguide width, considering a ±10% variation in the layer thicknesses of both SiN layers and of the oxide inter-layer.

Moreover, changes in the effective index of the optical mode lead to deviations in the hardwired GE emission phases, leading to a shift in the structured light sheet position relative to the targeted beam crossing area $S$. Figure 13(c) shows the change in the effective index of the TE ground mode for waveguide widths spanning a range of 300 nm to 1 µm, assuming each layer to vary by ±10% relative to its nominal thickness. While thinner waveguides have a refractive index that depends strongly on both biases in the layer thicknesses and the waveguide width, the sensitivity of wider waveguides beyond ~0.7 µm on waveguide width is strongly reduced, so that implementing the straight sections of the distribution network with multimode waveguides would reduce at least this process sensitivity. However, as we shall see in the following, the sensitivity of the GE emission angle on trench width and the sensitivity of the light sheet position on the waveguide width can also partially offset each other by keeping the light sheet aligned with *S*, in a similar manner as already observed for the wavelength sensitivities in Section 4.

We exemplify the effect of GE emission angle and waveguide effective index changes with two sets of simulations: First, we consider the hardwired phases at the 525 nm central wavelength to be implemented as targeted while varying the emission angle; second, we fix the emission angle to its design value and vary the effective index of the waveguides. The interplay between these two variations is discussed thereafter.

Figure 14(a) shows the intensity profiles resulting from deviations in the emission angle by ±5° relative to the nominal design, without corresponding adjustments of Eq. (7). These emission angle variations are much larger than what is expected from realistic process variations and are chosen to exemplify what happens in extreme cases. They result in the target position of the light sheet no longer coinciding with the area *S* where the beams emitted by the individual GEs cross. Consequently, two separate, distorted light sheets form, one at the position where the beams generated by the left side of the OPA intersect the targeted *x*-coordinate of the light sheet, as determined by Eq. (7), the other where the beams generated by the right side of the OPA intersect it. In both, the fringes resulting from the interference of the left and right sides of the OPA are no longer visible. The exaggerated ±5° target emission angle variation corresponds to a beam displacement of ±353 µm, which is significant relative to the MFD of the beams emitted by the GEs, that span ±237.5 µm.

Figure 14(b) shows how the shape of the light sheet is changed as the emission angle is changed from its nominal value by the maximum expected increment of 0.4° resulting from stacking the effect of layer thickness and trench width variations. It is apparent that the generated light sheets are very similar to each other and that this does not significantly degrade performance. This is because the beam crossing point *S* is displaced by only 28 µm, well below the width of the area *S*. However, this does reduce by the same amount the "misalignment budget" between beam crossing point and interference pattern position that is available for wavelength dependent beam steering, so that the latter would lead to some additional power translation degradation unless the steering range is reduced accordingly or the wavelength tuning range recentered.

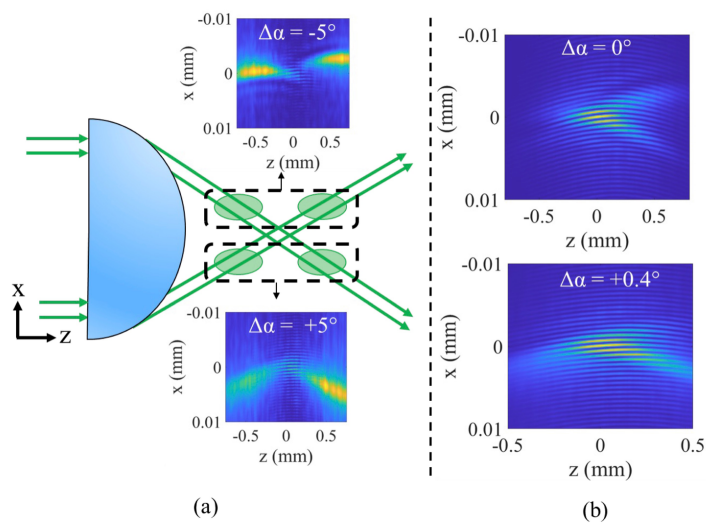

Fig. 14. (a) Illustration of the consequences of an off-target GE emission angle on the light sheet formation. The insets show the intensity profiles obtained for extreme deviations of the emission angle by $\Delta\alpha$ = -5° (top) and $\Delta\alpha$ = +5° (bottom). (b) Intensity profiles in the *xz*-plane (top view) obtained for the nominal emission angle and a maximum expected deviation +0.4° resulting from stacking up the sensitivities to trench width and layer thicknesses. In all cases, the x-axis of the graphs has been recentered on the actual beam crossing point.

Figure 15 shows the results of simulations in which the effective index of the waveguide has been varied by ±0.005. This results in a translation of the light sheet by ±30 μm along the lateral *x*-direction. Importantly, smaller waveguides, which correspond to an over-etch, result in a smaller effective index and in a displacement towards lower *x*-coordinates, which is the same direction in which the crossing point is also moved due to wider trenches in the GE. After empirically determining the statistical correlation between the process biases applied to isolated waveguides and to GE trenches, the width of the interconnect waveguides can be chosen with help of the data in Fig. 13 such that the two displacements are of equal magnitude and the performance of the system becomes immune to this process variation. To check that the two effects are in a range in which they can compensate each other, we can make the simplifying assumption that the waveguide and unetched grate width change by the same amount. The -0.4° emission angle change resulting from changing the trench width from 240 to 260 nm (and thus reducing the unetched grate width from 125 to 105 nm) results in a displacement of the beam crossing point by -28.8 μm that can be compensated by reducing the waveguide effective index by -0.005. This is for example achieved if the interconnect waveguide width is changed from 400 to 380 nm.

The effective index variation resulting from changes in the layer thicknesses, on the other hand, cannot be readily compensated by corresponding sensitivities in the GEs, that are of much smaller magnitude. These are significant and the ±0.017 effective index change seen in Fig. 13(c) due to the top SiN layer thickness change would lead to light sheet displacements of ±100 μm, which is 7× larger than the corresponding displacements in *S*. While still below half the individual Gaussian beam's MFD and thus manageable, this variability should ideally be guarded against by other means as it would otherwise substantially reduce the steering range. This can be achieved by having multiple devices on the same chip, each corresponding to an effective index increment of 0.005 in the evaluation of Eq. (7), and picking the best one post fabrication.

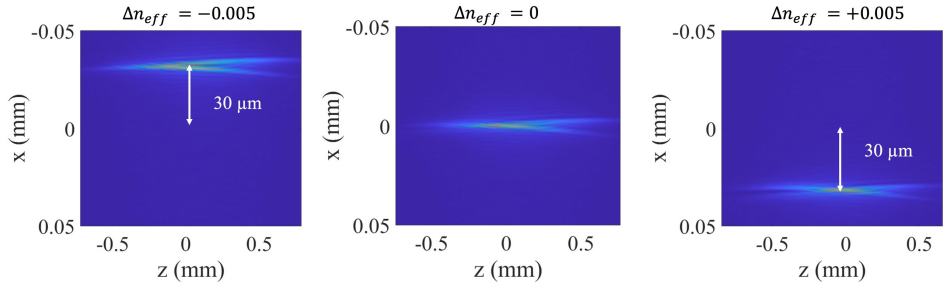

Fig. 15. Intensity profiles in the *xz*-plane (top view) obtained for waveguide effective index variations between -0.005 and +0.005. The corresponding light sheet displacements are between -30 μm and +30 μm.

## 6. Coherence length requirements

The topic of the previous section consists in systematic process biases applied across the entire OPA. Random and localized variations in the distribution network, however, can also lead to relative phase errors between GEs, i.e., the waveguide coherence length is an essential metric for the performance of the beam forming system. We follow the usual definition of the coherence length $L_{coh}$, wherein the phase error induced by roughness and random layer thickness variations in a waveguide of length *L* is assumed to have a standard deviation $\sigma_\varphi$ given by [49]

$$\sigma_\varphi = \sqrt{2L/L_{coh}} \qquad (8)$$

We further assume this phase error to follow a Gaussian distribution and determine it independently for each waveguide segment based on a random number generator that is independently run in each simulation.

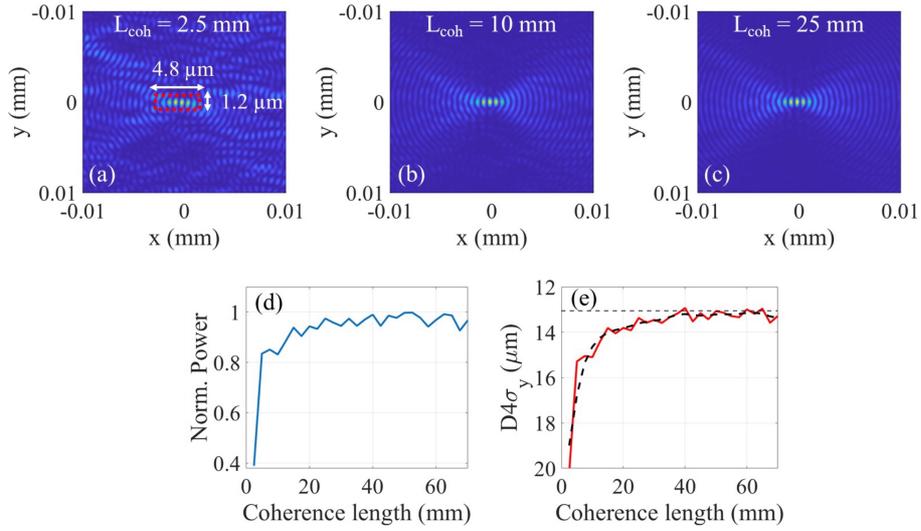

Fig. 16. (a)-(c) Intensity profiles in the front focal (*xy*-)plane of the illumination lens considering coherence lengths of 2.5, 10, and 25 mm, respectively. (d) Power in the central lobes of the structured light sheet resulting from integrating the intensity in a rectangular area with dimensions 4.8 by 1.2 µm as a function of the coherence length. This power is normalized to the one obtained in a reference simulation with infinite coherence length. (e) Light sheet thickness along the *y*-direction as evaluated with the D4σ method, also as a function of the coherence length. The *y*-axis is flipped in (e) to highlight the similarity between the two curves.

Figures 16(a)-(c) present intensity profiles simulated for coherence lengths of 2.5, 10, and 25 mm, in which the varying levels of deterioration relative to the ideal case (see Fig. 11(b)) can be seen. There is a clear qualitative difference between the three panels, with 25 mm displaying an almost ideal profile, some slight stray light visible for 10 mm coherence length, and very strong deterioration at 2.5 mm.

This deterioration can be quantified further with metrics such as the power contained at the center of the sheet, as displayed in Fig. 16(d). As in the previous sections, the intensity is integrated in a rectangular area whose width and height are taken as two times the corresponding FWHM of the ideal beam obtained with infinite coherence length. This power is further normalized to that of the reference simulation with infinite coherence length. A 20% power drop is observed at a coherence length of 5 mm. At 23 mm coherence length, which is a number in reach of state-of-the-art SiN technologies [49], this power penalty is reduced to 5%.

Another quantitative metric is given by the D4σ sheet thickness in the *y*-direction, calculated according to

$$\text{D4}\sigma_y = 4\sqrt{\frac{\iint (y-y_0)^2 \cdot I(x,y) \cdot dxdy}{\iint I(x,y) \cdot dxdy}} \quad (9)$$

where $y_0$ is the center position of the beam along the *y*-axis and $I(x,y)$ is the local intensity of the beam.

Here too, it can be observed that it asymptotically reaches the value of the reference simulation at large coherence lengths, with the corresponding curve in Fig. 16(e) taking almost the same shape as the power confinement shown in Fig. 16(d). The larger values at low coherence length are indicative of the stray light at large offsets in the *y*-direction, that effectively increases the beam width. It may be noted that even in the reference case, the D4σ sheet thickness is much larger than the FWHM, $w_y$. This is because contrary to Gaussian beams, the D4σ beam width is an undefined integral for an ideal Bessel-like beam and diverges as the integration domain becomes larger. Here, the integration domain was restricted to ±10 µm in the *y*-direction, which determines the asymptotic value of the curve.

The asymptotic value in itself is thus not of particular relevance, it is rather the trend over coherence length that is revealing.

## 7. Two-wavelength structure

Figure 17 illustrates the schematic of an envisioned dual-color SiN-PIC-based LSFM beam former designed to generate structured light sheets at two distinct central operating wavelengths of 525 nm and 637 nm. This PIC facilitates lateral beam scanning (along the *x*-axis) through wavelength tuning for both structured light sheets. Since the operating principle closely resembles that of a single-color PIC-based LSFM beam former, we will not reiterate previously discussed aspects here but will rather emphasize the architectural aspects that are tied to the dual-wavelength operation.

In addition to the excitation of the TdTomato fluorescent marker, which is excited by the beam at 525 nm, this PIC also facilitates imaging of a cleared sample stained with the fluorochrome smURFP. This fluorochrome efficiently absorbs and emits light at 642 nm and 670 nm, respectively. Similar considerations are made, such as maintaining a high excitation efficiency at the utilized wavelength (> 90% from peak), ensuring a sufficient distance from the peak emission wavelength to enable effective filtering, and ensuring compatibility with commercially available tunable lasers. These factors lead to the selection of an excitation laser operating at $637 \pm 5$ nm for the smURFP fluorescent marker.

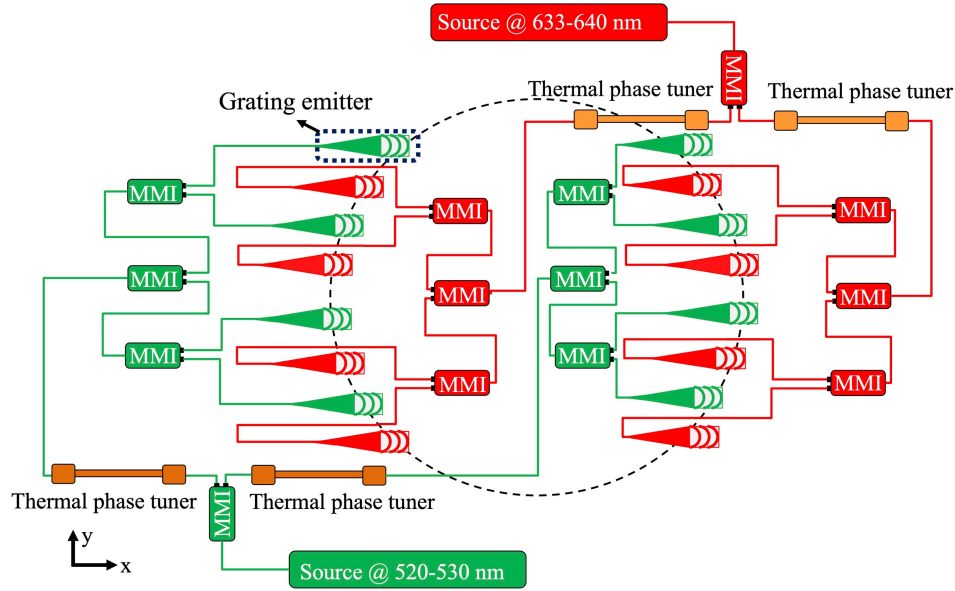

Fig. 17. PIC schematic of a dual-color LSFM beam former.

As a first challenge, the GE used for 637 nm needs to have an emission angle $\alpha_{637}$ that is close to that of the 525 nm GE, so that the GE-emission crossing-points for both are co-localized at the front focal plane of the lens. To facilitate this, the GEs for both wavelengths are also oriented in the same direction. To interleave the two networks, one is fed in from the left and the other from the right. However, the differential waveguide lengths should ideally also be proportional to +*x* for the forward oriented GEs of the 637 nm network. This ensures that the wavelength dependent emission angle and the wavelength dependent light sheet translation, as well as the etch bias induced emission angle change and the etch bias induced light sheet translation, continue to offset each other, as described for the 525 nm OPA in Sections 4 and 5. It is achieved by looping back the waveguides providing light to the 637 nm GEs with 180° bends at a common *x*-coordinate for all the waveguides on a given side of the OPA, see Fig. 17.

The 637 nm GE is designed with 250-nm-wide trenches and 205-nm-wide grates. This yields a forward emission angle of $\alpha_{637}$ also centered on 5.5° that decreases from 5.75° to

5.25° as the wavelength increases from 632 to 642 nm. These GEs are also non-apodised, consist of 35 trenches, and are designed to expand the 620-nm-wide input waveguide mode through a taper with a full angle of 22°, with the first grate placed 1 μm after the taper's onset. The emitted beam is best approximated by a Gaussian beam with an MFD of 10 μm, which is used in the raytracing simulations. Based on 3D FDTD simulations, 44% of the input power is extracted through the top side of the PIC, with a Gaussian beam overlap of 70%. We assume a total of 512 GEs for each wavelength, to be arranged on the circumference of a circle with $R = 3$ mm and an angular coverage of $\theta = 120°$ on both sides of the OPA, which leaves enough space to interleave the two types of grating couplers on the same circle (to total number of GEs is 1024 as in the layout depicted in Fig. 10(c)).

Figures 18(a)-18(f) show the simulated intensity profile of the structured light sheet generated at 637 nm. The light sheet thickness $w_y$ is determined to be 0.75 μm. The width $w_x$ and the collimation length $w_z$ are 3.2 μm and 509 μm, respectively.

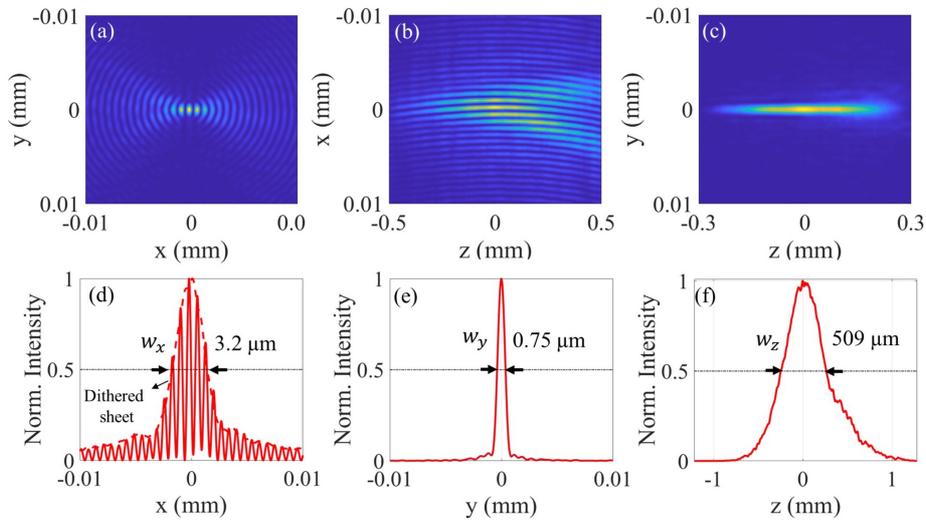

Fig. 18. Intensity distribution for the light sheet generated at 637 nm in (a) the *xy*-plane, (b) the *xz*-plane, (c) the *yz*-plane, and along (d) the *x*-axis, (e) the *y*-axis and (f) the *z*-axis. The curve in (d) is overlaid with the one obtained after applying dithering. The curve in (f) is also generated after applying dithering, to prevent the slanted orientation of the interference fringes to skew the data.

Figure 19 shows the beam steering obtained by tuning the operation wavelength in the range of 632 nm to 642 nm, resulting in ± 98 μm beam steering in the lateral direction. The areas that can be covered by both sheets are thus similarly sized and both matched to fit the FOV of a 100× objective.

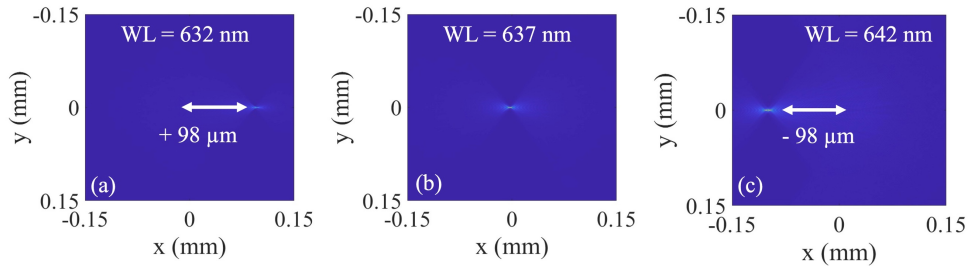

Fig. 19. Beam steering by wavelength tuning, with intensity profiles shown at (a) 632 nm, (b) 637 nm, and (c) 642 nm.

As a future step, the visible wavelengths external cavity lasers required for wavelength-based beam steering could also be integrated with the OPAs into a single SiN chip [50, 51].

## 8. Conclusions

We have modeled a beam shaping system for light sheet microscopy based on a PIC and an illumination lens, that allows compact, wavelength-based steering of the sheet without any moving parts. A sheet thickness of 0.6 μm (FWHM), that enables sub-micrometer axial resolution, is combined with a diffraction-less depth of 665 μm and a lateral steering range of ±98 μm obtained from a ±5 nm wavelength tuning range centered on 525 nm that stays within the absorption spectrum of the targeted fluorophores. A waveguide coherence length of 5 mm leads to the power contained in the main lobe of the generated interference pattern to be 80% of that obtained from an infinite one, with the 20% power penalty dropping to 5% for a 23 mm coherence length in reach of state-of-the-art fabrication processes. Finally, a concept is presented that allows the generation of a sheet with two excitation wavelengths, whose corresponding beams can both be steered across the lateral direction.

**Funding.** German Federal Ministry for Research and Education (BMBF, 13N15965), German Research Foundation (DFG, RTG 2610).